\documentclass[twocolumn,showpacs,amssymb, amsmath,nobibnotes, aps, prl]{revtex4}
\usepackage{graphicx}
\usepackage{bm}%
\usepackage[dvips]{color}

\begin{document}
\title{Wind instability of a foam layer sandwiched between the atmosphere and the ocean}
\author{Yuri M. Shtemler}
\email{shtemler@bgu.ac.il}
\author{Ephim Golbraikh}
\email{golbref@bgu.ac.il}
\author{Michael Mond}%
\email{mond@bgu.ac.il}\altaffiliation{\\} \affiliation{* \ddag
Department of Mechanical Engineering,
\dag Center for MHD studies,\\ Ben-Gurion University of the Negev, \\
P.O.Box 653, Beer-Sheva 84105,
Israel}%

\date{\today}

\begin{abstract}

The Kelvin-Helmholtz (KH) instability of short gravity waves is
examined in order to explain the recent findings of the decrease
in momentum transfer from hurricane winds to sea waves. A foam
layer between the atmosphere and the ocean is suggested to provide
significant stabilization of the sea-water surface by the
wavelength shift of the instability towards smaller scales. It is
conjectured that such stabilization leads to the observed drag
reduction. The problem of a three-fluid system with large
differences in densities provides an extension to the fundamental
KH problem in fluid mechanics.
\end{abstract}
\pacs{92.60.Cc, 92.10.Fj }
\maketitle
{ \it Introduction.---} Results of direct measurements
extrapolated from weak to strong winds predict a linear increase
of momentum transfer from wind to sea waves. The present study is
motivated by recent findings of saturation and even decrease in
the drag coefficient (capping) in hurricane conditions that is
accompanied by production of a foam layer on the ocean surface
\cite{powell}. A possible explanation for the
phenomenon is the development of a foam layer at the air-sea
interface. The principal role of such an air-water foam layer in
energy dissipation and momentum transfer from hurricane wind to
sea waves has been first suggested in \cite{newell}.
Winds generate waves on the ocean surface with a wide spectrum of
wave lengths. The longest waves, hundreds meters of length, attempt
to catch up with the wind, while the steeper short waves break out
and play a dominant role in drag production
\cite{donelan}-\cite{soloviev}.
 When the wind speed exceeds storm force (~24m/s),
wave breaking creates  streaks of bubbles near the ocean surface.
 As the wind exceeds the
hurricane force (~32m/s),
 streaks of bubbles combined with patches of foam cover
 the ocean surface.
When the wind speed   exceeds ~50m/s, a foam layer completely
covers the ocean surface \cite{powell}.

  Nowadays, there is a little hope for a comprehensive numerical calculations
  of the drag coefficient reduction that includes a detailed description
   of the wave breaking and foam layer production.
  Indeed, up to now there is no complete
understanding of the phenomenon.
In the present study, the intermediately short wave
Kelvin-Helmholtz instability (KHI) \cite{drazin}-\cite{alexakis}
of a foam layer between the atmosphere and the ocean is
investigated in order to qualitatively explain the drag reduction
phenomenon.
Such three-layer system exhibits a high contrast in densities of
constituting fluids $\rho_a\ll\rho_f\ll\rho_w$.
The present study is not concerned with the formation mechanism of
the foam layer by the hurricane but rather focuses on how a foam
layer isolates the lower atmosphere from the sea surface
 The existence of the foam layer on the ocean
surface is postulated and supported by observations (see
\cite{powell}, \cite{reul} and references therein).
The present modeling demonstrates a new effective mechanism to
stabilize the sea surface by a thin foam layer  between the
atmosphere and the ocean. However, beyond that particular
application, the current work addresses a fundamental problem in
fluid mechanics which provides a generalization of the classic
KHI. Thus, the peculiarities of three-layer systems with large
differences in the densities may be of interest to a wide range of
applications in the laboratory as well as in geophysics and
astrophysics.

{\it The physical model.---}
 A
piecewise constant approximation for the equilibrium densities and
for the longitudinal velocities of the water, foam and air
$\rho_j$ and $U_j$,  $(j=a,f,w)$ is employed:
\begin{eqnarray}
   \rho =\rho_w,    U=U_w\equiv 0 \,\,\,\, \,\mbox{for} \,\,\, \,  y<0,
 \nonumber
\\
\,\,\,\,\,\,\,\,\,\,\,\,\,\,\, \rho =\rho_f,    U=U_f  \,\,\,
\,\,\,\, \,\mbox{for} \,\,\,  0<y<L_f,
 \nonumber
\\
 \rho =\rho_a,    U=U_a   \,\,\, \,\mbox{for} \,\,\, \,
  y>0.
\end{eqnarray}
Here   $U_a$ is the known  constant  velocity of the wind, while
the constant foam layer thickness $L_f$ and velocity $U_f$ are the
widely unknown parameters of the foam layer in hurricane
conditions. In addition, it is assumed that the equilibrium state
is in hydrostatic equilibrium, namely,
  $\partial P_j/\partial y =-g \rho_j$ ($g$
is the gravity acceleration).

  The equations of motion that govern the
  dynamics of the system   in each of the three layers, and the
 appropriate boundary conditions
    are applied at the foam layer interfaces.
The equilibrium state is perturbed as follows:
\begin{eqnarray}
\Phi(x,y,z)=F(y)+F'(x,y,t),
\end{eqnarray}
where $\Phi$  stands for any of the physical variables, and $F$
and $F'$ denote  the equilibrium and perturbed values,
respectively. The latter are assumed to be of the form
$F'=f'(y)exp(-i\omega t+ikx)$ with real  $k$ and complex
$\omega=\omega_r+i\omega_i$. Thus, the amplitudes $f'$ that
satisfy the boundary conditions at $y=\pm\infty$ are given by:
\begin{eqnarray}
f'_a=\tilde{f}_a \exp(-ky),\,\,f'_w=\tilde{f}_w \exp(ky),
  \nonumber
  \\
f'_f=\tilde{f}_{-f}\exp(-k  y)+\tilde{f}_{+f}\exp(k y),
\end{eqnarray}
 where tilde denotes constant magnitudes.

Finally, capillary
and viscosity effects  are neglected for both the equilibrium
and  perturbed states (see the section {\it Results and
discussion}).
Substitution of Eqs. (1)-(3) into the linearized Euler equations
and applying the  continuity  conditions of normal velocity and
pressure at the foam interfaces, yields the
quartic dispersion relation for phase velocity $C$
\cite{craik}:
 \begin {eqnarray}
2(H_a +H_w )+(E-1)(H_a+1)(H_w+1)=0,
\end{eqnarray}
 where
 \begin{eqnarray}
C={\omega \over k},\,\,\,\,\,\,\,\, H_w= { \rho_w( U_w-C)^2-
(\rho_w-\rho_f) g/k\over \rho_f( U_f-C)^2},
 \nonumber
\\
E=exp(2kL_f ),
 H_a= { \rho_a( U_a-C)^2-( \rho_f-\rho_a) g/k\over \rho_f(
U_f-C)^2}.
\end{eqnarray}
%
Before turning to the  study of the foam layer effect, it is
noticed that in the limit $L_{f}=0$, or, equivalently, either
$\rho_{f}=\rho _{w}$ or $\rho_{f}=\rho _{a}$, Eq. (4) is reduced
to the classic dispersion relation $H_a+H_w=0$ for KHI
%
\cite{drazin}:
 \begin{equation}
\rho_w(k_0 U_w- \omega_0)^2+\rho_a(k_0 U_a- \omega_0)^2= k_0
g(\rho_w-\rho_a),
\end{equation}
where the subscript 0 denotes the foam-free parameters.

{\it Asymptotic analysis.---}
First, the limit of low air-water density ratio,
$\rho_a/\rho_w=\epsilon^2\ll 1$, ($\epsilon^2\approx10^{-3}$)
 is  applied to the classic two-layer case described by Eq. (6)
 with $U_w =0$, in order to obtain an estimate for the various physical parameters:
 \begin{eqnarray}
 \omega_0=\sqrt{gk_0-\epsilon^2 k_0^2 U_a^2}+O(\epsilon^2 k_0 U_a,
 \epsilon {gk_0}/\omega_0).
\end{eqnarray}
 Doing so, it can be concluded that the
 classic two-fluid KHI
 is excited in the short wavelength regime:
 \begin{eqnarray}
k_0 L_{*}\sim k_0^{*}L_{*}= {1 /\epsilon^{2}},\,\,\, {\omega_0
L_{*}/U_*}\sim{1 /\epsilon},\,\,C_0/ U_*\sim{\epsilon},
\end{eqnarray}
where
$U_{*}=U_a,\,\,\,\,L_{*}=U_a^2/g$,
while the superscript asterisk denotes the marginal values of the
parameters.

 Back to the general case of three-fluid
 systems, it is assumed that the
water content in the foam, $\alpha_w$, is small (low water content
is a characteristic feature of air-water foams).
As a result, $\alpha_w\sim 0.05$ is scaled with $\epsilon$ and
yields
\begin{eqnarray}
{\rho_f \over \rho_*}\approx  \alpha_w \sim \epsilon,\,\, \,
{\rho_a \over \rho_f}\approx {1\over \alpha_w}{\rho_a \over
\rho_*}\equiv {\epsilon^2\over\alpha_w}\sim \epsilon.
\end{eqnarray}
Here  $\rho_{*}=\rho_{w}$, $\rho_f=\alpha_a \rho_a+\alpha_w
\rho_w$,  $\alpha_a=1-\alpha_w$.
Assuming now that the three-fluid system operates in the same
regime
that gives rise to the KHI in the classic air-water system,
the following scales are adopted:
\begin{eqnarray}
kL_{*}\sim {1\over \epsilon^2},\,\, {\omega L_{*}\over U_*}\sim{1
\over\epsilon},\,\,  {C\over U_*}\sim{\epsilon}.
\end{eqnarray}
Further assuming that the foam layer thickness is much less than
the characteristic length, $L_f/L_{*}\ll 1$, ($L_{*}\sim 250m$ for
$U_a \sim 50m/s$), while the foam velocity is much less then the
wind velocity and much larger the phase velocity $\epsilon\sim
C/U_{*}\ll U_f/U_{*}\ll 1 $:
\begin{eqnarray}
{U_f/ U_*}\sim \epsilon^a,\,\,{L_f/ L_*}\sim
\epsilon^b,\,\,\,0<a<1,\,\,\,0<b,
\end{eqnarray}
 which yields the following estimates for
%
%
Eq. (4):
\begin{eqnarray}
H_a \sim H_w \sim \epsilon^{1-2a},\,\, E\sim exp(\epsilon^{b-2}).
\end{eqnarray}
Inserting the scaling (12) into Eq. (4), and applying
 the principle of the least degeneracy
\cite{vandyke} of the three-fluid problem,
 results in $a=1/2$, $b=2$, which means:
 \begin{equation}
{U_f\over U_*} \sim \epsilon^{1/2},\,\,
 {L_f\over L_*} \sim {{\lambda_0^*} \over {L_*}}\sim {1 \over
 {k_0^*L_*}}
  ={\rho_a \over \rho_* }\sim\epsilon^{2},
 \end{equation}
 where $\lambda_0^{*}=2 \pi /k_0^{*}$.
Following  relations (13), the wave number and frequency are
rescaled as follows:
%
\begin{equation}
\hat{k}={k/ k_0^*}\sim\epsilon^0,\,\, \,\,\hat{\omega}={\omega /
\sqrt{g k_0^*}}\sim\epsilon^0.
 \end{equation}
This yields the
dispersion relation to leading order in $\epsilon$:
\begin{equation}
\hat{\omega}=\sqrt{{
2(\hat{k}-\hat{k}^2)-({E}-1)(\hat{k}^2}K_f-\hat{k})
(K_f^{-1}+1)\over {2+({E}-1)(K_f^{-1}+1)}},
  \end{equation}
where $E=exp(2\hat{k}\hat{L}_f)$, while the rescaled foam
thickness $\hat{L}_f$, and the equilibrium ratio of the
foam-to-air dynamic pressure $K_f$ ($0<K_f<1$) are given by:
 \begin{equation}
\hat{L}_f=k_0^*L_f\sim \epsilon^0,\,\,\,K_f={\rho_f U_f^2 \over
\rho_a U_a^2}\sim \epsilon^0.
 \end{equation}
Thus, the system stability is parameterized by the dimensionless
foam velocity and thickness or, equivalently, $K_f$ and
$k_0^*L_f$,
which  has a meaning of a bulk foam Richardson number $Ri_f$
scaled by $\rho_a/\rho_f=\epsilon^2/\alpha_w\sim\epsilon$:
 \begin{eqnarray}
 \nonumber
 \hat{Ri}_f=k_0^*L_f,\,\,\,Ri_f=-g{\Delta\rho
\over \rho_f}{ L_f \over \Delta U^2}\approx {\rho_a\over \rho_f}
\hat{Ri}_f,
\end{eqnarray}
 where  $\Delta U=U_a-U_w\equiv U_*$ and
$\Delta\rho=\rho_a-\rho_w\approx-\rho_*$.
  %

Two particular limits  of Eq. (15) are readily obtained, namely,
the foam-free limit ($H_w+H_a=0$ for $L_f=0$):
 \begin{equation}
 {\omega_0 \over \sqrt{g k_0^*}}=i\sqrt{ {k^2\over k_0^{*2}} -{k\over
 k_0^{*}}},\,\,  \hat{R}i_f=k_0^{*}L_f=0.
 \end{equation}
and the foam-saturated  limit ($H_w+1=0$ for $L_f=\infty$):
\begin{equation}
 {\omega_\infty \over
  \sqrt{g k_\infty^*}}
 =i\sqrt{ {k^2\over k_\infty^{*2}} -{k\over
 k_\infty^{*}}},\,\,\hat{R}i_f=k_0^{*}L_f=\infty,
 \end{equation}
 which differs from Eq. (17) by replacing $k_0^*$,  $\omega_0$
with $k_\infty^* =k_0^*/K_f$,  $\omega_\infty $ ($0<K_f<1$).
 Comparison of these two limits demonstrates the
stabilizing effect of the foam  due to the decrease of the
marginal wavelength from the foam free
$\lambda_0^* =2\pi /k_0^*$ to
 the foam-saturated
$\lambda_\infty^* =2\pi /k_\infty^*$  value.
 The growth rate $\omega_i$ decreases
 %
%
  from the foam-free $\omega_{i0}$ to the
foam-saturated $\omega_{i\infty}$ value.
%
%
The definition for $k_\infty^* =k_0^*/K_f$  {\color{black}is used}
in order to express $K_f$ through $\lambda_\infty^*$: $K_f=
k_0^*/k_\infty^*\equiv\lambda_\infty^*/(2\pi \epsilon^2L_*)$.
{\color{black}The intermediate wavelength value $\lambda\approx
\lambda_\infty^*\approx 1 m$ is chosen for further estimations
from the wavelength range of the drag responsible waves $\sim
0.1-10 m$.
In turn, a typical height  $h \approx 0.1m$ is expressed from
Stokes heuristic rule for the critical steepness of breaking waves
\cite{fringer}.
}
%
%
 Consequently, the value
 $K_f\approx 0.5$ is adopted
%
%
that results in
 $U_f=\epsilon
U_a\sqrt{K_f/\alpha_w}\approx 5m/s$ at $\alpha_w\approx 0.05$,
$U_a\approx 50m/s$.
\begin{figure}[!h]
\includegraphics[scale=0.25]{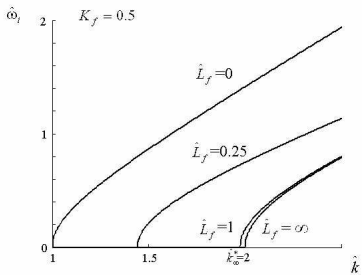}
\caption{\label{fig:epsfig1}
{\color{black}
Growth rate $\hat{\omega}_i=\omega_i/\sqrt{gk_0^*}$  vs  wave
number, $\hat{k}=k/k_0^*$, for the typical foam-layer thicknesses,
$\hat{R}i_f\equiv\hat{L}_f=k_0^*L_f$ and the ratio of the foam/air
dynamic pressure $K_f=0.5$.
 }
 }
 \end{figure}
Figure ~\ref{fig:epsfig1} depicts the growth rate
as a function of the
wavenumber.
As can be seen, the growth rate decreases
as the foam layer thickness is increased and approaches
%
%
its saturated limit
already at
$k_0^*L_f\approx 1$.
\begin{figure}[!h]
\includegraphics[scale=0.25]{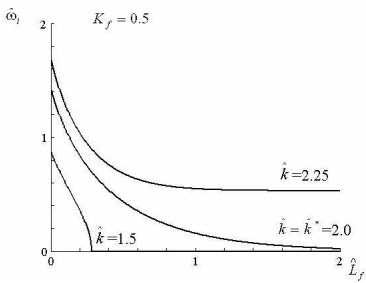}
\caption{\label{fig:epsfig2}
{\color{black}
Growth rate $\hat{\omega}_i=\omega_i/\sqrt{gk_0^*}$  vs foam-layer
thickness, $\hat{R}i_f\equiv\hat{L}_f=k_0^*L_f$,
 for the typical wave number $\hat{k}=k/k_0^*$  and the
ratio of the foam/air dynamic pressure $K_f=0.5$.
 }
 }
\end{figure}
The dependence of the growth rate $\omega_i/\sqrt{gk_0^*}$ on the
foam-layer thicknesses
is depicted in Fig.~\ref{fig:epsfig2}. For
sufficiently short waves ($k/k_0^*>1/K_f$) the growth rate
strongly  drops from the foam free value at
$k_0^*L_f$=0 to its saturation level at
 foam-layer thickness $k_0^*L_f\approx 1$.
 The growth rates of perturbations with longer waves ($k/k_0^*<1/K_f$)
 sharply decrease with the increase
 $k_0^*L_f$,
till total stabilization at a finite value of $k_0^*L_f$ is
achieved. These two cases are separated by the threshold curve
($k/k_0^*=1/K_f$) for which the growth rate vanishes at
$k_0^*L_f>>1$.
%

The  marginal wave number $k^*$ satisfies the eigenvalue equation
for the three-layer system:
\begin{equation}
exp(2k^*L_f)=1-{2\over 1+K_f^{-1}}{1-k^*/k^*_0 \over 1-
k^*/k^*_\infty}.
\end{equation}
As in the  classic two-fluid system, to leading order in
$\epsilon$, the waves propagate with phase velocity $C=\omega/k$
without amplification for $k/k^*<1$, and amplify with zero phase
velocity for  $k/k^*>1$.
The
value
$k^*$ monotonically increases with $k_0^*L_f$ from
the foam-free value
$k^*=k_0^*$
%
%
to the foam-saturated value
$k^*=k_\infty^*\equiv k_0^*/K_f$.
%

{\it Results and discussion.---} The atmosphere-ocean interaction
in hurricane conditions
%
%
  {\color {black}creates}    a foam layer
between the atmosphere and the ocean. This provides for an
effective mechanism of the sea surface stabilization.
%
%

The analysis of the KHI is treated asymptotically in two small
parameters: air-water density ratio $\sim \epsilon^2$ and
 water content in the
foam $\sim \epsilon$.
The system stability is parameterized by the dimensionless foam
 velocity $U_f$ and thickness $L_f$ (or, equivalently,
 the dynamic pressure
ratio $K_f$ and   Richardson number $\hat{R}i_f$).
 Due
to lack of observations or modelling data in hurricane
%
environment, they are  first estimated as
 $L_f/L_*=\epsilon^2$ and $U_f/U_*
\sim\sqrt{\epsilon}$ by applying the asymptotic principle of least
degeneracy of the problem.
 Then
$L_f^{(ef)}\approx 0.25m$ at $U_*\approx 50m/s$ is evaluated by
the condition that the growth rate approaches its minimal
saturated value at
$L_f^{(ef)}$ {\color{black} $=\epsilon^2L_*$},
and further increase
 $L_f$ is ineffective, as if the foam layer is
of infinite thickness. The value   $L_f^{(ef)}$ is of the order of
the experimentally observed values (\cite{reul} and references
therein).
 The single fitting parameter of
the model $K_f=
\lambda_\infty^*/\lambda_0^* \approx0.5$ ($U_f\approx 5m/s$) had
been estimated  {\color{black} through an intermediate value of
length of drag responsible waves ($\lambda\approx1m$)}.
The value of the wavelength ratio exhibits the  instability shift
towards smaller wavelength scales. Thus, the foam layer reduces
the foam-free wavelength $\lambda_0^*$ approximately by a factor 2
to the foam saturated limit $\lambda_\infty^*$ already at
$L_f\approx L_f^{(ef)}$.
%
%
This scale-down in the characteristic unstable length scales
provides a qualitative link between the linear stability modeling
and the role of the foam layer in the air-sea momentum exchange.
%
%
%
%
To see that, the {\color{black} local correlation, based on the
dimensional grounds,} $z/\lambda=F(h/\lambda)$ between the ocean
surface roughness $z/\lambda$ and the wave steepness $h/\lambda$,
 is examined {\color{black}in a vicinity of
intermediate values of height and length of drag responsible
breaking waves.
It is similar to the correlation
 \cite{taylor} for pre-hurricane conditions, but with the local
 values
 $h, \lambda$ instead of
significant wave height and peak wavelength.}
 %
Noting that
 the breaking process does not completely destroy the waves,
but rather tears off their tops, when their steepness exceeds a
critical value (
%
%
determined by nonlinear effects),
%
%
the complete covering of the ocean surface by the foam layer
occurs when the critical steepness $1/10$
\cite{fringer} is achieved for drag responsible breaking waves. As
a result,
%
%
 the roughness is reduced along with the wave
length by  {\color{black} a factor $\lambda_0^*/\lambda_\infty^*=
K_f^{-1}\approx 2$ (for $\lambda_\infty^*\approx \lambda\approx
1m$, $h\approx 0.1m$, $K_f\approx0.5$)} due to the foam effect.
Remarkably, such simple scalings are supported by the observations
of the
%
%
 roughness and drag reduction presented in
\cite{powell}. The results are physically transparent, since in
the foam saturated system the foam layer totally separates the air
flow from the sea surface, and the three-fluid system becomes
close to a  two-fluid foam-water system. Formally this corresponds
to substituting the foam density and velocity instead of those
parameters for the air in the classic
 two-fluid model.

Finally, the main assumptions adopted in the present
 study are discussed.
First, it is noted that the assumption of foam thickness
uniformity, breaks down due to the foam accumulation in troughs of
high and long ocean waves
(seen in a photograph of the sea surface
 before its complete coverage by a foam \cite{powell}). The
drag of the hurricane induced surface waves comes mainly from
intermediately short waves (of $\sim 0.1-10 m$ \cite{chen}). Thus,
although the breaking waves under consideration (of $\sim 1m$) are
strongly modulated by
long waves {\color{black} (of $\sim 10^2m$)},
 such a slow variation of the foam thickness may be
taken into account in the next approximation. It is assumed that
the air-sea exchange of the highly solvable carbon dioxide or
oxygen leaves the equilibrium water content to be small in
compliance with the determining feature of gas-liquid foams. Zero
compressibility and viscosity approximation is commonly accepted
in KHI  of  air-water systems \cite{drazin}-\cite{alexakis}. The
foam compressibility may be ignored within the same accuracy as
the air one.
 Indeed, using the smallness of
 the air Mach
number $M_a=U_a/C_a$ and noting that the foam-to-air sound
velocity ratio $C_f/C_a\sim \sqrt{\rho _a/\rho _f}\sim
\sqrt{\epsilon }$ \cite{shtemler} is of the same order as
$U_f/U_a\sim \sqrt{\epsilon }$ (see Eqs. (13)), it is obtained
that $M_f=U_f/C_f\sim M_a\ll 1$.
 Although the
foam viscosity data in hurricane environment is unavailable,
artificial foam viscosities are known to be significantly larger
than the viscosity of its liquid and gas constituents. On the
other hand, natural sea foams are expected to have lower viscosity
than their artificial counterpart due to lack of man-made
surfactants
 and a larger effective size of the foam bubbles
(of$\sim 0.2-2mm$ \cite{soloviev}).
In any case, the stability behavior regarding the growth rate of
the shear viscosity (see e.g. \cite{bhata}) ignored in the present
study can only enhance the foam stabilizing effects in the range
of the intermediately short waves.
  Ignoring the capillary effects  is
valid at the water-foam interface since the foam is composed of
the same sea-water. At the air-foam interface the value of the
surface-tension coefficient may be naturally assumed equal to the
surface tension between air and sea-water.
 Since the latter is
 much less than the
surface tension between air and fresh-water due to the effect of
the surfactants, its  influence on KHI is  rather small for the
short waves
 (of $\sim 1m$).
 The bubbly liquid,  spray and foam  coexist in
hurricane environment.  The high  contrast in three-fluid
densities is the principal feature of the
 foam-layer system
 $\rho_a\ll\rho_f\ll\rho_w$, distinguishing it from the layers of bubbly liquid
or spray, with
  $\rho_b\approx \rho_w$
or $\rho_s\sim \rho_a$, when
the system will be  close to the two-fluid air-water
configuration.
 Indeed, the bubbly liquid
superposed the  water
 hard to be distinguished due to
a small contrast in  densities ($\rho_b\approx\rho_w$, since
$\alpha_w\approx1$), and the bubbly liquid layer may be dropped
from the KHI study.
The spray between the air and the foam  is well distinguished from
them by a  contrast in densities.
Indeed, typically $\rho_f\approx\alpha_w\rho_w$, and
$\rho_s=\alpha_w\rho_w+\alpha_a\rho_a\approx
2\rho_a$ at $\alpha_w\approx0.001$ for spray and
$\alpha_w\approx0.05$ for foam, and $\rho_s\approx
2\rho_a\ll\rho_f$.
 Thus, the spray may be described
by a more detailed velocity and density  air profiles within
the present air-foam-water configuration. This
however will increase the system uncertainty,
 since the spray layer thickness and velocity are
rather unknown.
\acknowledgments Helpful discussions with  A. Soloviev and V.
Chernyavski
%
are gratefully acknowledged.

\end{document}